\documentclass[aps,pra,twocolumn,superscriptaddress,showpacs,amsmath,amssymb, amsfonts,10pt]{revtex4-1}

\usepackage{graphicx, placeins}
\usepackage[rgb]{xcolor}
\usepackage{amsmath,tikz}
\usepackage{booktabs}
\definecolor{darkred}{rgb}{0.7098039215686275, 0.09411764705882353, 0}
\usepackage[unicode=true,pdfusetitle,bookmarks=true,bookmarksnumbered=true,bookmarksopen=false, breaklinks=false,pdfborder={0 0 0},pdfborderstyle={},backref=false,colorlinks,bookmarks=false,urlcolor=darkred, citecolor=darkred, linkcolor = darkred, pdftitle={DrivenDissipativeLRO}]{hyperref}
\usepackage{lipsum}  
\usepackage{color}
\usepackage{verbatim}
\usepackage[T1]{fontenc}

\usepackage[normalem]{ulem} 
\usepackage{braket}
\usepackage{appendix}

\usepackage{cancel}
\usepackage{color}
\newcommand{\al}{\alpha}
\newcommand{\bb}{\beta}

\newcommand{\xv}{\mathbf{x}}

\newcommand{\rr}{\mathbf{r}}

\newcommand{\qq}{\mathbf{q}}
\newcommand{\cH}{\mathcal{H}}

\begin{document}
\title{ 
Stabilization of long-range order in low-dimensional nonequilibrium $O(N)$ models
}
\author{Oriana K. Diessel}
\email[E-Mail: ]{oriana.diessel@cfa.harvard.edu}
\affiliation{ITAMP, Center for Astrophysics, Harvard \& Smithsonian, Cambridge, Massachusetts 02138, USA}
\affiliation{Department of Physics, Harvard University, Cambridge, Massachusetts 02138, USA}
\author{Jaewon Kim}
\affiliation{Department of Physics, University of California, Berkeley, CA 94720, USA}
\author{Ehud Altman}
\affiliation{Department of Physics, University of California, Berkeley, CA 94720, USA}

\date{\today}
\begin{abstract}
It is now well established that the Mermin-Wagner theorem can be circumvented in nonequilibrium systems, allowing for the spontaneous breaking of a continuous symmetry and the emergence of long-range order in low dimensions. However, only a few models demonstrating this violation are known, and they often rely on specific mechanisms that may not be generally applicable.
In this work, we identify a new mechanism for nonequilibrium-induced long-range order in a class of $O(N)$-symmetric models. Inspired by the role of long-range spatial interactions in equilibrium, consider the effect of non-Markovian dissipation, in stabilizing long range order in low-dimensional nonequilibrium systems. We find that this alone is insufficient, but the interplay of non-Markovian dissipation and slow modes due to conservation laws can effectively suppress fluctuations and stabilize  long-range order.
\end{abstract}

\maketitle

\textit{Introduction.---}
Non-equilibrium systems sometimes show striking emergent phenomena that cannot exist in thermal equilibrium. In particular, it has been shown that driven systems can exhibit spontaneous breaking of continuous symmetries in two dimensions, which would be prohibited in thermal equilibrium by the Mermin-Wagner theorem~\cite{Mermin_1966,Hohenberg_1967}. The mechanisms already known to enable such violations include self convection of moving particles in the ordering of two dimensional "flocks" \cite{vicsek1995novel,Toner1995}, convection induced by externally imposed shear flow\cite{corberi2003correlation}, and anisotropic relaxation in models with a conserved order parameter \cite{Bassler_1995}.

In this paper we describe a novel mechanism for stabilizing spontaneous breaking of $O(N)$ symmetry in low-dimensional non-equilibrium systems. The mechanism operates via the interplay of two essential elements: coupling to a non-Markovian bath and conserved symmetry charges. The presence of these conserved charges gives rise to slow diffusive modes that play a crucial role in our mechanism. This element is present, for example, in the dynamics of an
$O(3)$ antiferromagnetic order parameter subject to spin conservation or in the dynamics of a superfluid ($O(2)$ order parameter) subject to charge conservation.

\begin{figure}[t]
	\includegraphics[width=1
	\linewidth]{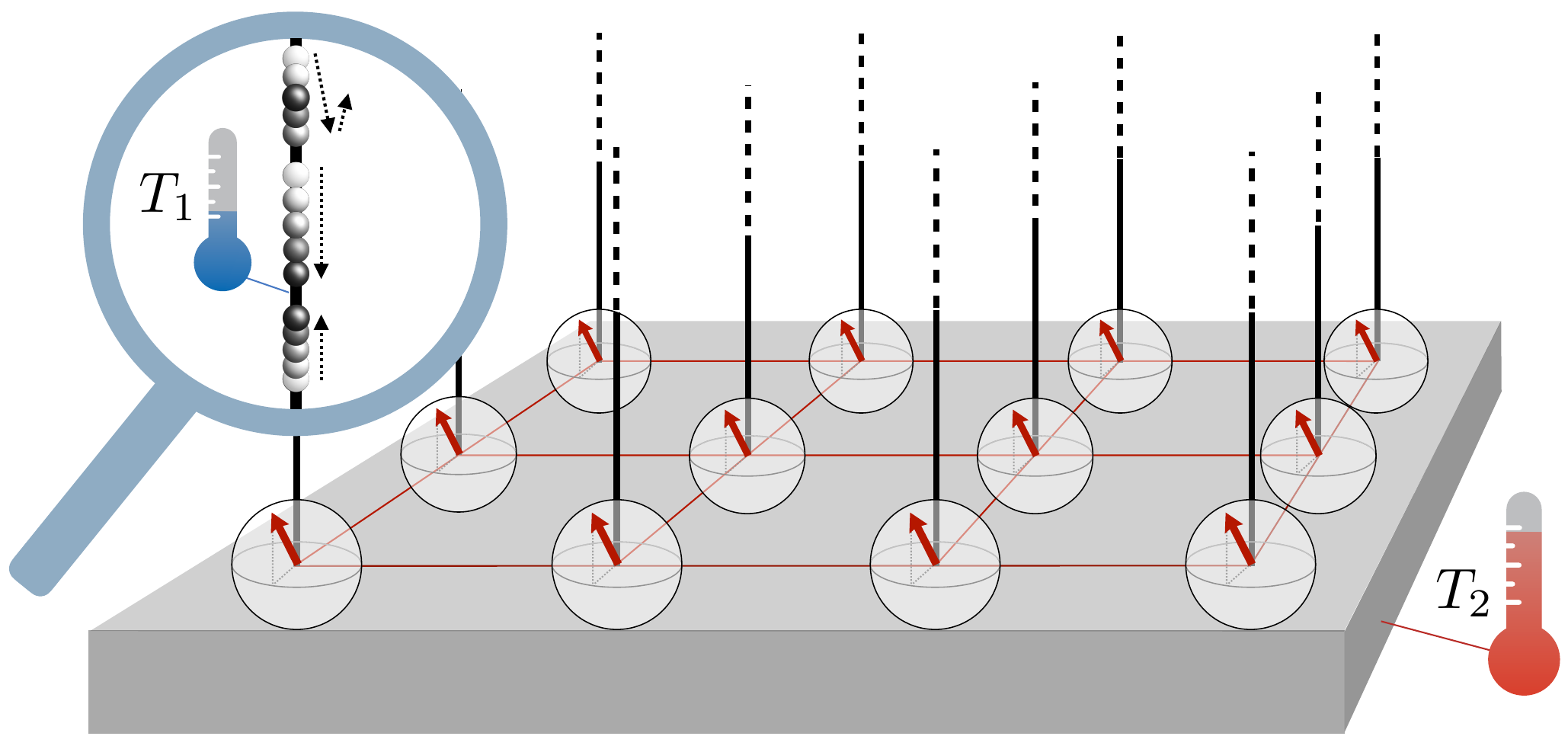}
 \caption{Schematics of a system that is coupled to a non-markovian bath: 
 A two-dimensional array of $O(3)$ rotors is coupled to one-dimensional wires hosting diffusing particles (see inset). This coupling generates both a non-markovian dissipation kernel and non-markovian noise correlations in the effective dynamics of the rotors. To prevent the rotors from thermalizing with the bath, they are coupled to another bath at a different temperature $T_1\neq T_2$ (gray substrate).}
 \label{fig:MainFig}
\end{figure}

Non-Markovian dissipation emerges naturally when the system couples to baths with a broad-tailed distribution of relaxation times. The concept is illustrated with a physical example in Fig.~\ref{fig:MainFig}: a two-dimensional system pierced by an array of perpendicular wires hosting diffusing particles. The particles in the wires, maintained at temperature $T_1$,  interact with the system at the intersection points. This interaction generates both a dissipation kernel and noise correlations in the effective dynamics of the 2D system that decay as power laws in time, reflecting the algebraic decay of the diffusive autocorrelation functions in the wire baths. 
Note, however, that if the system were coupled only to this bath, then it would thermalize with the bath. Consequently, the order parameter correlations would approach equilibrium values, and no violations of the Mermin-Wagner theorem would occur~\cite{Bonart_2012,aron2010symmetries,Schmalian_2014}. 
To prevent this thermal fate the system must couple to an additional bath maintained at a different temperature $T_2\ne T_1$, as depicted by the slab surrounding the 2D system in Fig. \ref{fig:MainFig}. This second bath can be Markovian.

The proposition that non-Markovian dissipation can help stabilize long range order draws on an analogy with equilibrium systems. Long-range temporal interactions induced by the dissipative memory kernel could plausibly have a similar effect as long-range spatial interactions, which effectively stabilize long-range order in equilibrium systems~\cite{fisher1972,sak1973}. However we find below that the memory kernels cannot do this alone. Coupling to the two-dimensional diffusive modes of the conserved symmetry charges is required for translating the long temporal memory, provided by the non-markovian bath, to effective long-range spatial interactions.

{\it Model. --} As a concrete model, we study the dynamics of an $O(N)$ order parameter field $n_\alpha$ and its corresponding angular momenta $L_{\alpha\beta}\equiv \epsilon_{\alpha\beta}\,n_\alpha \pi_\beta$, where $\pi_\alpha$ are the  momenta conjugate to $n_\alpha$ , i.e., $\{n_\alpha, \pi_\beta \} = g\, \delta_{\alpha\beta}$, with $g$ a real number. 
Specifically we consider the dynamics generated by the following Langevin equations:
\begin{subequations}
\label{eq:Langevin}
    \begin{align}
        \gamma_n*\dot{n}_\alpha &= \{n_\alpha,\cH\}-\frac{\delta \cH}{\delta n_\alpha}+ \xi_\alpha \,,
        \label{eq:Langevin_n} \\
        \gamma_L* \dot{L}_{\alpha\beta} &= \{L_{\alpha\beta},\cH\}-(i\nabla)^{a}\frac{\delta \cH}{\delta L_{\beta\alpha}}+ \eta_{\alpha\beta} \,.
        \label{eq:Langevin_L}
    \end{align}
\end{subequations}
Here the $*$ products denote convolutions with the non-Markovian memory kernels $\gamma_{n,L}(t)$.  $\cH$ is the Hamiltonian function
\begin{equation}
    \cH=\frac{1}{2}\int_\rr \left(D_LL^2_{\al\bb}+D_n(\nabla n_{\al})^2 + r n_\alpha^2 + \frac{u}{2}n_\beta^2 n_\alpha^2 \right).
    \label{Eq:hamiltonian}
\end{equation}
$\xi_\alpha$ and $\eta_{\alpha \beta}$ are zero-average Gaussian noise fields with correlation functions given by:
\begin{subequations}
\label{eq:noise_correlations}
\begin{align}
    &\langle \xi_\alpha(\rr,t) \xi_\beta(\mathbf{0},0) \rangle =T\delta_{\alpha\beta} \delta(\rr) \gamma^s_n(t), \\
    &  \langle \eta_{\alpha\beta}(\rr,t) \eta_{\gamma\delta}(\mathbf{0},0) \rangle = T \Delta_{\alpha \beta\gamma\delta}(i\nabla)^{a}\delta(\rr)\gamma^s_L(t), 
\end{align}
\end{subequations}
where $\Delta_{\alpha \beta\gamma\delta}\equiv\delta_{\alpha\gamma}\delta_{\beta\delta}-\delta_{\alpha\delta}\delta_{\beta\gamma}$ and where we have defined the symmetrized memory kernels $\gamma_{n,L}^s(t) \equiv \gamma_{n,L}(t)+\gamma_{n,L}(-t)$. 
These equations describe dynamics that conserve the total angular momentum for $a=2$, whereas for $a=0$ angular momentum is not conserved.  

We take the memory kernels to have a power-law form $\gamma_{i}(t) = \theta(t) / t^{\alpha_{i}} \,, \ i = n,L,$ with $\theta$ the Heaviside function.
These kernels can be regarded as the causal response functions of baths characterized by power-law decay of temporal correlations~\cite{Bonart_2012}. It is most natural if these kernels originate from the same bath which couples through a neutral operator to both the order parameter and angular momenta fields. In this case we would expect $
\alpha_n=\alpha_L$, however we consider also the more general case of two different exponents.

The case  $\alpha_i=1$ represents an Ohmic environment, while $\alpha_i<1$ is sub-Ohmic~\cite{Weiss_2012}.
In the Super-Ohmic case $\alpha_i>1$ the Fourier transform gives rise to white noise at low frequencies, therefore the system becomes effectively Markovian in this case (see App.~\ref{App:RelevanceNonMarkovian}).
It is important to note that the memory kernels in general have both a dissipative and reactive (coherent) component, where the former is odd under time-reversal, while the latter is even. In Apps.~\ref{App:NonEq_EffTemp},~\ref{App:ProofNonEq} we prove that the steady state of this model cannot be described by a thermal equilibrium (Gibbs) distribution. We therefore argue that these equations provide a phenomenological description of non-equilibrium systems, such as depicted in Fig. \ref{fig:MainFig}.

Finally we note that if the dissipation kernels are taken to have a short memory, e.g.  $\gamma_{n,L}\propto \delta(t)$, then this dynamics belongs  to the usual classification of dynamical critical phenomena in thermal equilibrium \cite{Hohenberg_1977}. In particular, for an $O(3)$ order parameter and $a=2$ the markovian limit is identical to model $G$, describing the dynamics of an antiferrmagnet with conserved spin, while for $O(2)$ it  describes the dissipative dynamics of a superfluid order parameter with conserved charge.

{\it Linear fluctuation analysis. --}
To investigate the stability of long range order we examine the linearized Langevin equations describing the Goldstone modes of the putative ordered phase. Using a scaling analysis of the corresponding Gaussian fixed point we will determine if the Green's functions of the fluctuations exhibit infrared divergences that destabilize the order. 

We begin by parameterizing the fluctuations of the $O(N)$ order parameter about a uniform broken symmetry state as
\begin{multline}
    n_\alpha(\rr, t) = \left( \delta_{\alpha 1} \sqrt{\bar{n} -\varphi_\beta^2(\rr,t)} + (1-\delta_{\alpha 1}) \varphi_\alpha(\rr,t) \right)\\
    \times (1 + \sigma(\rr,t))
  \label{Eq:OrderparameterDecomposition}
\end{multline}
where $\bar{n}$ is the value of the order parameter in the symmetry-broken phase, $\sigma$ is the amplitude mode and $\varphi_\alpha$, with $\alpha = 2,\dots, N$, represent the $N-1$ Goldstone modes.
The value of $\bar{n}$ determined at the mean-field level is $\bar{n} = \sqrt{-r/u}$ for $r<0$. We also denote by $\ell_\alpha=L_{1\alpha}$, the angular momentum component involving the mean field order parameter $\bar{n}$.

The linearized Langevin equations are now given by

\begin{figure}[t]
	\includegraphics[width=1
	\linewidth]{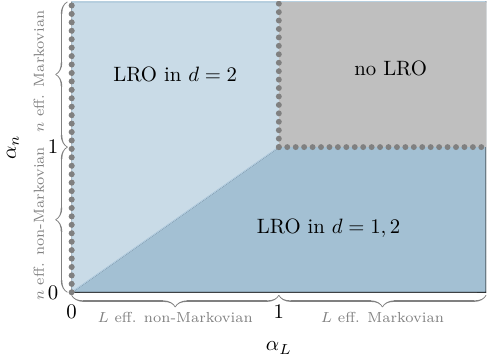}
 \caption{Phase diagram of the model~(\ref{eq:Langevin},\ref{Eq:hamiltonian},\ref{eq:noise_correlations}) for $a=1$ as a function of $\alpha_n$ and $\alpha_L$. For $\alpha_n,\alpha_L\geq 1$the system behaves effectively as a Markovian system and exhibits no long-range order.
 For $\alpha_L=0$, the $L_{\alpha\beta}$ field acquires a mass, leading to the absence of long-range order in this regime (indicated by the gray dotted line).}
  \label{Fig:Phasediagram}
\end{figure}

\begin{subequations}
\label{Eq:linearized_equations}
   \begin{align}
       \gamma_n * \dot{\sigma} & = -2 u \bar{n}^2 \sigma + D_n\nabla^2\sigma + \xi_1, \label{Eq:linearized1}\\
       \gamma_n* \dot{\varphi}_\alpha & = D_n\nabla^2\varphi_\alpha - g\bar{n}D_L \ell_{\alpha} + \xi_\alpha,\label{Eq:linearized2}  \\
       \gamma_L*\dot{\ell}_{\alpha} & = -D_L(i\nabla)^{a} \ell_{\alpha}  - g\bar{n}D_n\nabla^2\varphi_\alpha +\eta_{\alpha1}, \label{Eq:linearized3}  \\
 \gamma_L*\dot{L}_{\alpha\beta} & = -D_L(i\nabla)^{a} L_{\alpha\beta}   +\eta_{\alpha\beta}\label{Eq:linearized4}, 
   \end{align}
\end{subequations}

 At this point we can solve the linearized equations \eqref{Eq:linearized_equations} to obtain the equal time correlation function of the Goldstone mode in the stationary state and extract its long wavelength behavior
$C_{\varphi\varphi}(\qq) = \langle \varphi_\alpha(\qq,t)\varphi_\alpha(-\qq,t) \rangle\sim 1/q^{\eta_\varphi}$. In equilibrium $\eta_\varphi= 2$ leading to infrared divergences that destroy the long range order in one and two dimensions.  

In the non-equilibrium system we consider, the direct solution of the equations \eqref{Eq:linearized_equations} gives a numerical result for $C_{\varphi\varphi}(\qq)$. However we can also find an analytic solution for $\eta_\varphi$ as a function of $\alpha_L$ and $\alpha_n$ through a scaling analysis. Specifically, $\eta_\varphi$ is related to the scaling dimension of the field $\varphi_\alpha$ at the Gaussian fixed point describing the Goldstone modes in the putative broken symmetry state.

To evaluate the scaling dimensions of the different fields it is useful to recast the Langevin equations as a MSR action 
\begin{align}
   S_0 =\!\!\int_{\xv,t}\!\!\bigg\{&
   \tilde{\varphi}_{\al} \Big[\gamma_n * \dot{\varphi}_{\al}-D_n \nabla^2 \varphi_{\al} + g\bar{n}D_L\ell_\alpha - T\gamma^s_n*\tilde{\varphi}_{\al}\Big]
   \nonumber\\
  & +\tilde{\ell}_\alpha
   \Big[
   \gamma_L  * \dot{\ell}_\alpha + D_L(i\nabla)^{a}\ell_\alpha + g\bar{n}D_n\nabla^{2}\varphi_\alpha \nonumber\\
   & - T\gamma^s_L(t)*(i\nabla)^{a}\tilde{\ell}_\alpha \Big]
  \! \bigg\}.
  \label{eq:gaussian_action}
\end{align}
The fields $\tilde{\varphi}_\alpha$ and $\tilde{\ell}_\alpha$ are the usual response fields conjugate to $\varphi_\alpha$ 
and $\ell_\alpha$ respectively. 

Before proceeding, note that if $\alpha_L=1$ or $\alpha_n=1$ then the scaling of the corresponding convolution terms is identical to the Markovian case. Furthermore if $\alpha_i>1$, then the corresponding convolution term converges at long time and the term also reverts to markovian scaling. Hence if any $\alpha_i>1$ we should set it to $\alpha_i=1$ to obtain the correct scaling.  We now turn to the scaling analysis. In the model with conservation of angular momentum ($a=2$), the "diffusion" terms $D_n$ and $D_L$ cannot be both marginal. Indeed we find that they are irrelevant at the Gaussian fixed point of the non-Markovian model. The scaling dimensions of the fields and the dynamical exponent $z$ are then assigned so that all other terms are marginal. The result is shown in table ~\ref{tab:scaling}, where under rescaling of length $x\to x/b$, $\varphi_\alpha\to b^{[\varphi]}\varphi_\alpha$ (i.e. $[x]=-1$). 
The exponent $\eta_\varphi$ governing the low $\qq$ singularity of $C_{\varphi\varphi}(\qq)$ follows from a Fourier transform and is given by
\begin{equation}
\label{eq:eta_exponent}
 \eta_\varphi = \frac{2\alpha_n}{\alpha_n + \alpha_L}.   
\end{equation}
In App.~\ref{app:Plot}, we compare this analytical result to a numerical solution of the linearized equations \eqref{Eq:linearized_equations}, showing excellent agreement.

We see that in the entire range where the system is non Markovian (i.e. $\alpha_n<1$ and/or $\alpha_L<1$) we have $\eta_\varphi<2$ suggesting that LRO is stable in this regime in two dimensions. Furthermore for $\alpha_n<\alpha_L$ we even have $\eta_\varphi<1$ implying long range order in one dimension. The phase diagram inferred from the linearized fluctuation analysis is shown in  Fig.~\ref{Fig:Phasediagram}. In the next section, we confirm that non-linearities do not alter this conclusion.

\begin{table}[t]
    \centering
    \begin{tabular}{lcc}
        \toprule
        & $L$ conserving$\quad$ & $L$ non-conserving \\
        \midrule
        $[\tilde\varphi]$ $\quad$    & $(2-\alpha_n)/(\alpha_n+\alpha_L) + d/2$ & $2/\alpha_n + d/2 - 1 $ \\[1mm]
        $[\varphi]$           & $-\alpha_n/(\alpha_n+\alpha_L) + d/2$  & $d/2-1$ \\[1mm]
        $[\tilde{\ell}]$  & $(2-2\alpha_L-\alpha_n)/(\alpha_n + \alpha_L) + d/2$ & $2/\alpha_n +d/2 - 1$ \\[1mm]
        $[\ell]$          & $\alpha_n/(\alpha_n + \alpha_L) + d/2$ & $d/2+1$ \\[1mm]
        $\phantom{[}z$               & $2/(\alpha_n + \alpha_L)$ & $2/\alpha_n$ \\
        \bottomrule
    \end{tabular}
        \caption{Scaling dimensions of the four Goldstone fields and angular momenta $\tilde\varphi_\alpha, \ \varphi_\alpha, \ \tilde\ell_\alpha, \ \ell_\alpha$, and the dynamical exponent $z$.}
    \label{tab:scaling}
\end{table}

\begin{figure}[t]
	\includegraphics[width=1
	\linewidth]{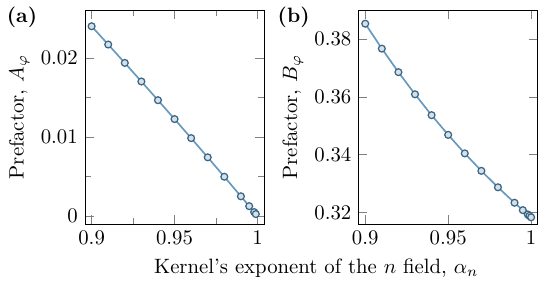}
 \caption{Dependence of the prefactor $A_{\varphi}$ [(a)] and $B_{\varphi}$ [(b)] on $\alpha_n$ for $\alpha_L=1$, when parametrizing the inverse correlation function for the $\varphi$-fields as $C_{\varphi\varphi}\approx 1/(A_{\varphi}|\qq|^{2\alpha_n/(\alpha_n+\alpha_L)}+B_{\varphi}q^2)$ for $\qq\rightarrow 0$.
As $\alpha_n$ approaches 1, $A_{\varphi}$ vanishes which leads to the expected scaling of $C_{\varphi\varphi}\approx 1/q^2$ in the markovian case.}
  \label{Fig:PrefactorPlot}
\end{figure}

In the Markovian regime we expect the usual equilibrium behavior with $C_{\varphi\varphi}(\qq) \approx 1/q^2$. Interestingly, the exponent $\eta_\varphi$ given in Eq. \eqref{eq:eta_exponent} does not approach $2$ in the limit $\alpha_{n}=\alpha_L=1$, where the system is expected to become Markovian. Instead, the scaling $1/q^2$ originates from a sub-leading term that becomes dominant at this point. Specifically, the correlation function in the long wavelength limit beyond the leading $q$ dependence is given by  $C_{\varphi\varphi}(\qq) \approx 1/(A_{\varphi} q^{2\alpha_n/(\alpha_n+\alpha_L)}+B_{\varphi} q^2)$. We can determine the prefactors $A_{\varphi}$ and $B_{\varphi}$ by fitting to the numerical solution of the linearized equations \eqref{Eq:linearized_equations}. 
The fitted values of those pre-factors are given in Fig.~\ref{Fig:PrefactorPlot} as a function of $\alpha_n$ for $\alpha_L=1$, showing that $A$ vanishes as $\alpha_n$ approaches 1, while $B_\varphi$ remains finite.

It is also interesting to understand the limiting cases in which either $\alpha_{n} = 0$ or $\alpha_{L} = 0$. In this case, the corresponding field develops an effective mass (see App. \ref{app:alphazero}), similar to Goldstone modes in presence of long-range interactions~\cite{Diessel_2023}. 
Therefore, along the line $\alpha_n=0$ and $\alpha_L \neq 0$, the Goldstone modes become gapped, there are no infrared divergencies and LRO is preserved. 
If, instead, $\alpha_L=0$ and $\alpha_n \neq 0$, the conserved density $L$ develops a gap and consequently decouples from the order parameter fluctuations in the long wavelength limit. The fixed point is then described by standard model $A$ dynamics of a non-conserved order parameter~\cite{Hohenberg_1977}. In this case $\eta_\varphi=2$, leading to the usual infrared divergence that destroys LRO in both one and two dimensions.

Finally we comment on the model with non-conserved angular momentum ($a=0$). In this case the scaling dimension of $\varphi_\alpha$ implies   $\eta_\varphi= 2$, which leads to the usual IR divergencies in both one and two dimensions. Hence  we must have angular momentum conservation in addition to coupling to the non Markovian bath to stabilize long range order. 

{\it Nonlinearities. --}
The linear stability analysis carried out above suggests that LRO can be stabilized in two dimensions in a non-Markovian $O(N)$ model with conservation of angular momentum. We now verify that this order is not destabilized by non-linearities that were ignored so far.

The leading nonlinearities in the Langevin equations~\eqref{eq:Langevin} are described by the following perturbations to the MSR action  in Eq.~\eqref{eq:gaussian_action}:

\begin{align}
\label{eq:Interacting_action}
   S_\text{int}=\int_{\rr,t}\!\bigg\{ \tilde{\varphi}_\alpha \!\bigg[
   gL_{\al\bb}\varphi_{\bb} 
-g'\tilde{L}_{\alpha\beta}\Big[\varphi_{\al}\nabla^2 \varphi_{\bb}-\varphi_{\bb}\nabla^2 \varphi_{\al}
  \Big]
    \!\bigg\}.
\end{align}
We note that the fields $L_{\alpha\beta}$ and $\tilde{L}_{\alpha\beta}$ have the same scaling dimensions as $\ell$ and $\tilde \ell$ respectively. This can be seen by promoting~\eqref{Eq:linearized4} to a Gaussian MSR action.

Now using the scaling dimensions of the fields from table~\ref{tab:scaling} we can directly evaluate the scaling dimensions of the couplings $g$ and $g'$, which are summarized in  table~\ref{tab:gg'u_scaling}.

We find that in the model with charge conservation the nonlinearities are always irrelevant in two dimensions and for $\alpha_n<\alpha_L$ in one dimension. Hence the non-linearities are not expected to destabilized the LRO where it is allowed by the linear fluctuation analysis. 
In the model with no charge conservation the nonlinearities are marginal in two dimension. In the special case of $O(2)$ symmetry this term translates to the Kardar-Parisi-Zhang non-linearity, which is indeed marginal in two dimensions. 
\begin{table}[h!]
    \centering
    \begin{tabular}{lcc}
        \toprule
        & $\quad L$ conserving$\quad$ & $\quad L$ non-conserving \\
        \midrule
        $[g]\quad$           &  $\alpha_n/(\alpha_n+\alpha_L) - d/2$ & $1-d/2$ \\[1mm]
        $[g']\quad$  & $\alpha_n/(\alpha_n+\alpha_L) - d/2$ & $1-d/2$ \\
        \bottomrule
    \end{tabular}
    \caption{Scaling dimensions of the non-linear terms $g$ and $g'$. In the charge conserving case, especially for $\alpha_n < \alpha_L$, the non-linearities are always irrelevant in two dimensions and for $\alpha_n<\alpha_L$ in one dimension, suggesting that the Gaussian fixed point is stable. In the charge non-conserving case, the non-linearities are marginal in two dimensions.}
    \label{tab:gg'u_scaling}
\end{table}

Another effect of the nonlinearities is that they can generate Markovian dissipation terms even if there were none at the outset (see App. \ref{App:GeneratingMarkovianPerturbations}). However, as we show in appendix \ref{App:RelevanceNonMarkovian}, these terms are less relevant than the original non-Markovian dissipation.

{\it Conclusions.--}
We have established the conditions for the stabilization of long-range order in one- and two-dimensional driven $O(N)$ models coupled to non-Markovian baths. Long range order, which is ruled out in equilibrium, is stabilized by the interplay of the conserved angular momenta with the long memory of the bath, decaying as a  power-law in time. If the angular momentum conjugate to the order parameter is not conserved, then the long range order is generally unstable.

\textit{Acknowledgments---} We gratefully acknowledge discussions with Alessio Chiocchetta and Sam Garratt. This work was supported in part by a Simons Investigator Award (E.A.). O.K.D acknowledges support from the NSF through a grant for ITAMP at Harvard University. Initial stages of this research were supported by the Deutsche Forschungsgemeinschaft under Germany's Excellence Strategy EXC 2181/1 - 390900948 (the Heidelberg STRUCTURES Excellence Cluster), and by the International Max Planck Research School for Quantum Science and Technology (IMPRS - QST).

\bibliography{biblio}

\appendix

\widetext


\section{Generation of markovian perturbations}
\label{App:GeneratingMarkovianPerturbations}

In this Appendix, we assess how the presence of interactions can generate markovian perturbations in the ordered phase. To this extent, we include non-linearities back in Eqs.~\eqref{Eq:linearized_equations} and neglect the equation for the amplitude mode. Note that a more precise treatment would be to eliminate $\sigma$ adiabatically, which would lead to new interactions and renormalization of existing interactions. This is however not relevant, as we only want to show that a markovian term can be generated.  
The equations then read $(\alpha,\beta = 2, \dots,N)$:
\begin{subequations}
   \begin{align}
       \gamma_n* \dot{\varphi}_\alpha & = D_n\nabla^2\varphi_\alpha - g\bar{n} D_L\ell_{\alpha}  -g D_L L_{\alpha\beta}\varphi_\beta +\xi_\alpha,\\
       \gamma_L*\dot{\ell}_{\alpha} & = D_L\nabla^2 L_{1\alpha}  - g\bar{n}D_n\nabla^2\varphi_\alpha +\eta_{\alpha1}, \\
       \gamma_L*\dot{L}_{\alpha\beta} & = D_L\nabla^2 L_{\alpha\beta} + g D_n(\varphi_\alpha\nabla^2\varphi_\beta - \varphi_\beta \nabla^2\varphi_\alpha)  +\eta_{\alpha\beta}.
   \end{align}
\end{subequations}
Within this approximation, the only non-linear coupling is between $\varphi_\alpha$ and $L_{\alpha\beta}$, while $\ell_{\alpha}$ is only linearly coupled to $\varphi_\alpha$. 
In order to show that markovian contributions can be generated by interactions, we focus only on $\varphi_\alpha$ and $L_{\alpha\beta}$, and neglect $L_{1\alpha}$ for now.
We can then write down the corresponding MSR action, which reads: 
\begin{multline}
   S=\int_{\rr,t}\!\bigg\{
   \tilde{\varphi}_{\al} \Big[\gamma_n*\dot{\varphi}_{\al}- D_n \nabla^2 \varphi_{\al} + gD_LL_{\al\bb}\varphi_{\bb} -T \tilde{n}_{\al}\Big] \\
+\tilde{L}_{\bb\al}\Big[\gamma_L*\dot{L}_{\al\bb}-gD_n\big\{\!(\nabla^2 \varphi_{\bb})\varphi_{\al}-(\nabla^2 \varphi_{\al})\varphi_{\bb}\!\big\}
   -
   D_L\nabla^2 L_{\al\bb}- T\nabla^2\tilde{L}_{\al\bb}\Big]
    \!\bigg\}\!\label{Eq:FullAction}.
\end{multline}
We now compute the one-loop correction to the retarded self-energy of the field $\varphi_\alpha$.

\begin{figure}[h!]
	\centering
	\includegraphics[width=0.88\linewidth]{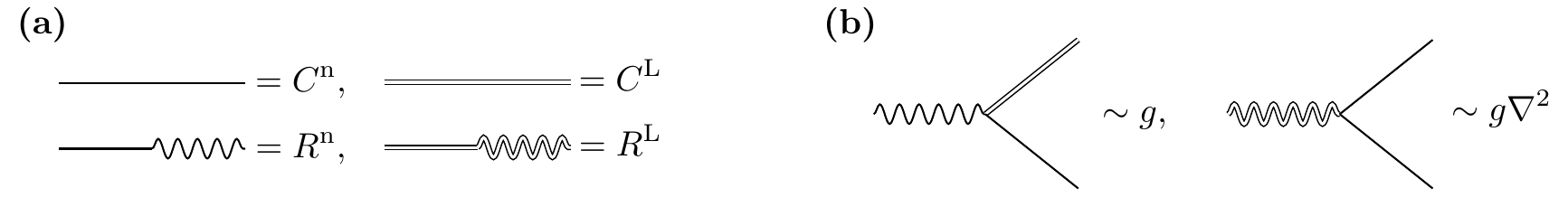}
    \label{Fig:FeynmanRulesApp}
    \caption{Definition of Feynman rules for \textbf{(a)} the correlation (C) and response (R) functions, and \textbf{(b)} the vertices.}
    \label{Fig:App2}
\end{figure}

To this end, we split the retarded self-energy in two parts as $\Sigma_R=\Sigma_R^{(1)}+\Sigma_R^{(2)}$, with:
\begin{figure}[h!]
	\centering
	\includegraphics[width=0.7
	\linewidth]{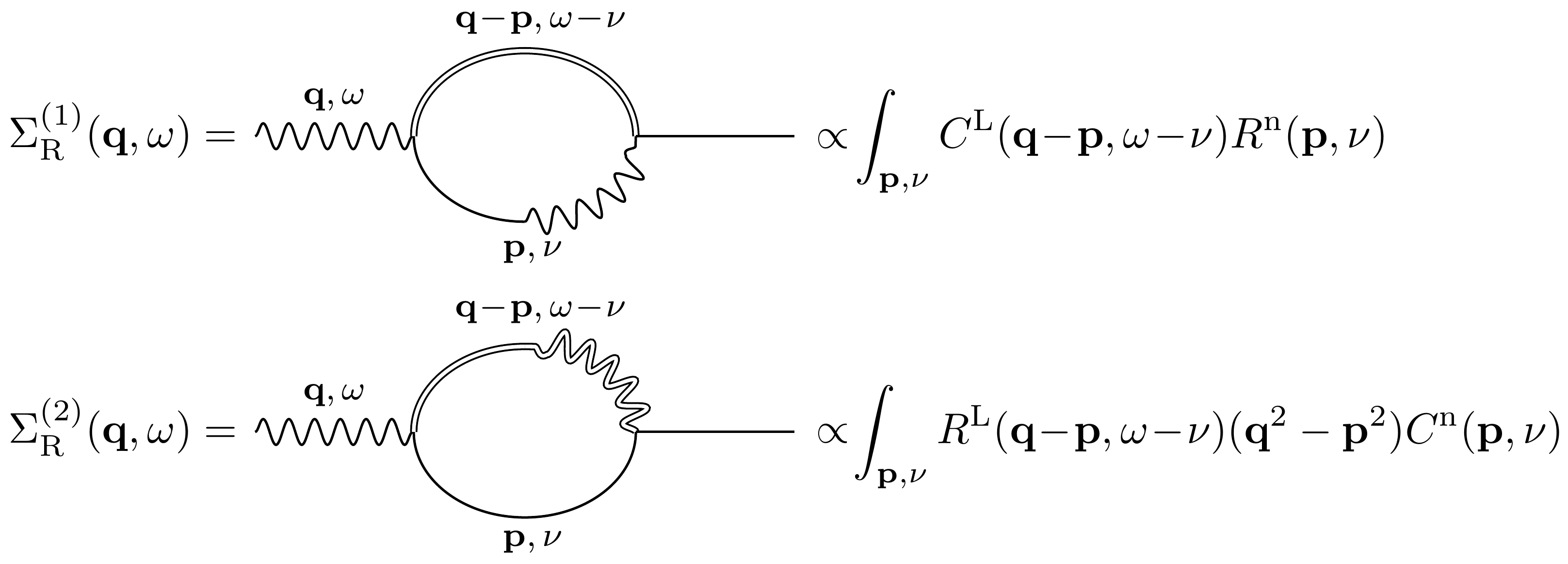}
\end{figure}

\noindent with the Feynman rules reported in Fig.~\ref{Fig:App2} and the Gaussian response ($R$) and correlation functions ($C$) are given by
\begin{subequations}
\label{eq:Gaussian_propagator}
\begin{align}
    C^L(\mathbf{q},\omega) & =\frac{2T_L \text{Re}[\gamma_L(\omega)]\mathbf{q}^2}{|-i\gamma_L(\omega)\omega + D_Lq^2|^2}\qquad\qquad
    R^L(\mathbf{q},\omega)=\frac{1}{-i\gamma_L(\omega)\omega+D_L\mathbf{q}^2}\\
    C^n(\mathbf{q},\omega) & =\frac{2T_n \text{Re}[\gamma_n(\omega)]}{|-i\gamma_n(\omega)\omega +D_n\mathbf{q}^2|^2} \qquad\qquad
    R^n(\mathbf{q},\omega)=\frac{1}{-i\gamma_n(\omega)\omega+D_n\mathbf{q}^2}.
\end{align}
\end{subequations}
In order to single out the generated markovian contribution $-i\gamma_\text{mkv}\omega$ we need to compute:
\begin{align}
\gamma_\text{mkv}
& =  \frac{\partial }{\partial\omega}\text{Im}[\Sigma_R (\omega, \mathbf{q})]\bigg|_{\omega = 0, \mathbf{q} = 0} \\
& =  g^2 \text{Im}\bigg[\int_{\mathbf{p},\nu} \left[ \frac{\partial C^L(\mathbf{p},\nu)}{\partial \nu}R^n(\mathbf{p},\nu) -\mathbf{p}^2 \frac{\partial R^L(\mathbf{p},\nu)}{\partial \nu}C^n(\mathbf{p},\nu) \right]\bigg] \\
&=g^2 \int_{\mathbf{p},\nu} \left[ \frac{\partial C^L(\mathbf{p},\nu)}{\partial \nu}\text{Im}[R^n(\mathbf{p},\nu)] -\mathbf{p}^2 \frac{\partial \text{Im}[R^L(\mathbf{p},\nu)]}{\partial \nu}C^n(\mathbf{p},\nu) \right]\bigg] 
\end{align}
Now, assuming this contribution to be evaluated within an RG step (say, Wilson's momentum shell scheme), the correlations and response functions would be those of fast degrees of freedom living within the momentum shell: accordingly, we can replace $\mathbf{p} \approx \Lambda$, with $\Lambda$ the UV cutoff, and the computation is simplified as:
\begin{equation}
    \gamma_\text{mkv} = d\Lambda \Lambda^{d-1} g^2 \int_{\nu} \left[ \frac{\partial C^L(\Lambda,\nu)}{\partial \nu}\text{Im}[R^n(\Lambda,\nu)] -\Lambda^2 \frac{\partial \text{Im}[R^L(\Lambda,\nu)]}{\partial \nu}C^n(\Lambda,\nu) \right] 
\end{equation}
with $d\Lambda$ the width of the momentum shell. By inserting in the previous equation the values~\eqref{eq:Gaussian_propagator}, together with the expression for $\gamma(\omega)$ in Eq.~\eqref{eq:gamma_expression}, one finds that the integral in $\nu$ converges, proving that markovian corrections are indeed generated.


\section{Scaling of non-markovian versus markovian dynamics}
\label{App:RelevanceNonMarkovian}
The presence of interactions can generate markovian perturbations in the ordered phase, as shown in App.~\ref{App:GeneratingMarkovianPerturbations}. In this appendix, we therefore investigate the stability of the non-markovian dynamics against markovian perturbation. To this end, we consider the case in which $\gamma(t)$ also has a markovian component, i.e., 
\begin{equation}
    \gamma(t) =  \gamma\frac{\theta(t)}{t^\alpha}\, f\left(\frac{t}{t_\Lambda}\right) + \gamma_\text{mkv}\,\delta(t),
\end{equation}
which appears as a delta function in time. Note also that we added the regulator $f(x)$ with the short-time cutoff $t_\Lambda$ in order to contemplate also the case $\alpha>1$, for which one has an UV (i.e., short time) divergence when taking the Fourier transform.
To assess the relevance of the different terms, it is convenient to take its Fourier transform, which reads (for $\omega t_\Lambda \ll 1$):
\begin{equation}
\label{eq:gamma_expression}
    \gamma(\omega) = \gamma \left[ \Gamma(1-\alpha) (-i\omega)^{\alpha-1} + \frac{t_\Lambda^{1-\alpha}}{\alpha-1} + O(\omega) \right] + \gamma_\text{mkv},
\end{equation}
with $\Gamma$ the gamma function. 
From~\eqref{eq:gamma_expression}, it is straightforward to identify three different regimes:

\textit{Sub-ohmic dissipation ($\alpha < 1$):} In this regime, the memory kernel reads $\gamma(\omega) = \gamma \Gamma(1-\alpha) (-i\omega)^{\alpha-1} + O(\omega^0)$. The non-markovian term dominates over the markovian one at low frequencies. Therefore, any markovian perturbation is subleading. 

\textit{Ohmic dissipation ($\alpha = 1$):} Here, the memory kernel reads $\gamma(\omega) = -\gamma \log(\omega t_\Lambda) + O(\omega^0)$. The non-markovian term gives a logarithmic correction to the markovian one.

\textit{Super-ohmic dissipation ($\alpha > 1$):} In this regime, the memory kernel reads $\gamma(\omega) = \gamma \frac{t_\Lambda^{1-\alpha}}{\alpha-1} + \gamma_\text{mkv} + O[\min(\omega,\omega^{\alpha-1} )]$.
The leading term is the Markovian component, making the system effectively Markovian at low frequencies. Notably, even in the absence of an explicit Markovian perturbation (i.e., $\gamma_\text{mkv} =0$), a Markovian-like term is generated through the regularization 
$t_\Lambda$.


\section{Proof of non-equilibrium conditions from fluctuation dissipation relations}
\label{App:NonEq_EffTemp}

In order to test whether the model defined at the outset describes thermal equilibrium conditions we can directly examine  the validity of the fluctuation-dissipation Theorem. In equilibrium the relation between the correlation function and the susceptibility should be 
\begin{equation}
C(\omega, \qq) = \frac{T}{2\omega} \text{Im}\chi(\omega, \qq)    
\end{equation}
where $C(\omega, \qq)$ is the correlation function in momentum and frequency, and $\chi(\omega, \qq)$ is the susceptibility to an external perturbation coupled to the order parameter.

For the purpose of this calculation we focus on the linearized Langevin equations, where both the correlation and response functions can be calculated explicitly. In the next Appendix we apply a more comprehensive test of equilibrium conditions, valid beyond the linearized model.
To quantify the deviation from thermal equilbirium  we introduce an effective temperature $T_\text{eff}(\omega, \qq)$ ~\cite{cugliandolo1997energy,cugliandolo2011effective} defined by the ratio
\begin{equation}
    T_\text{eff}(\omega, \qq) \equiv 2\omega \frac{C(\omega, \qq)}{\text{Im}\chi(\omega, \qq)},
\end{equation}
Under equilibrium conditions $T_{\text{eff}}$ is a constant independent of $\qq$ and $\omega$ and defines the actual temperature in thermal equilibrium. Even if the system is not strictly in equilibrium one may identify emergent equilibrium behavior of the slow modes ifthe effective temperature approaches a constant in the low frequency and long wavelength limit. 
On the other hand, if  $T_\text{eff}(\omega, \qq=0) \propto \omega^s$, then the low frequency modes do not have a well defined temperature and are therefore intrinsically non thermal.
%
%
The starting point for our analysis are the Linearized equations for the Goldstone modes
\begin{subequations}
\label{eq:Langevin__}
    \begin{align}
        \gamma_n*\dot{\varphi}_\alpha &= \{\varphi_\alpha,H\}-\frac{\delta H}{\delta \varphi_\alpha}+ \xi_\alpha \,,
        \label{eq:Langevin__n} \\
        \gamma_L* \dot{\ell}_{\alpha} & = \{\ell_{\alpha},H\}+\nabla^{a}\frac{\delta H}{\delta \ell_{\alpha}}+ \eta_{\alpha} ,
        \label{eq:Langevin__L}
    \end{align}
\end{subequations}
where the Hamiltonian $H$ is given as,
\begin{equation}
    H=\frac{1}{2}\int_\rr \bigg[D_L\ell^2_{\al}+D_n(\nabla \varphi_{\al})^2 \bigg].
    \label{Eq:Hamiltonian}
\end{equation}
Here $\varphi_\alpha$ are the Goldstone modes, $\ell_\alpha \equiv L_{\alpha1}$ are the conserved densities for $a=2$, or a massive field for $a=0$. Notice that $\varphi_\alpha$ and $\ell_\alpha$ are conjugate variables because $\{\varphi_\alpha, \ell_\beta\} = g\bar{n} \delta_{\alpha\beta} $, with $\bar{n}$ the mean-field value of the order parameter. The noise correlations are 
    \begin{align}
        &\langle \xi_\alpha(\rr,t) \xi_\beta(\mathbf{0},0) \rangle =T\delta_{\alpha\beta} \delta(\rr) [\gamma_n(t) + \gamma_n(-t) ], \label{eq:noise_correlations__}\\
        &\langle \eta_{\alpha}(\rr,t) \eta_{\beta}(\mathbf{0},0) \rangle = T \delta_{\alpha\beta}  (i\nabla)^a\delta(\rr)[\gamma_L(t) + \gamma_L(-t) ], \nonumber
    \end{align}

To compute the susceoptibility $\chi$ we add an external field to the Hamiltonian, so $H \to H - \int_\rr h_\alpha\varphi_\alpha $,  and obtain the susceptibility as the ratio :
\begin{equation}
    \chi_\alpha(\rr,t-t') = \frac{\delta \langle \varphi_\alpha(\rr, t)\rangle_h }{\delta h(\rr=0,t')}\Bigg|_{h =0}.
\end{equation}
Notice that $\chi$ is different from the response function in this case because the external field couples also to the conserved density via the Poisson bracket~\cite{tauber2014critical}.
Evaluating the correlation function and susceptibility, we find:
\begin{equation}
   T_\text{eff}(\omega, \qq) =  \frac{T}{1 + \frac{(g\bar{n})^2 \left[q^a\text{Re}(\gamma_n - \gamma_L) + 
    \omega \text{Im}(\gamma_L \gamma_n) \right]}{
 (q^{2 a}+\omega^2 |\gamma_L|^2 )\text{Re}(\gamma_n) + 
  q^a [2 \omega \text{Im}(\gamma_L)\text{Re} (\gamma_n) + (g\bar{n})^2 \text{Re}(\gamma_L)] 
  }}
\end{equation}

We start by checking the Markovian case. Assuming $\gamma_n(\omega) = \gamma_L(\omega) \equiv \gamma$ constant, it is straightforward to see that $T_\text{eff}(\omega, \qq) =T$, i.e., the system is at thermal equilibrium, as expected.
It is also instructive to consider the markovian case in which $\gamma_n$ and $\gamma_L$ are constant but different values. In this case the system is out of equilibrium, as it can be be mapped (by properly rescaling fields and time) onto a system with two different temperatures $T_n$ and $T_L$. The effective temperature is accordingly frequency- and momentum-dependent, but at low energy and momenta it reads:
\begin{equation}
    T_\text{eff}(\omega\to0, \qq\to0) = T \left(\frac{\gamma_L}{\gamma_n}\right)^{\frac{2-a}{2}}
\end{equation}
i.e., is a constant, regardless of the conserved ($a=2$) or not ($a=0$) dynamics. Therfore, we expect thermal behavior at low frequencies. 
We now consider the non-Markovian case. By using the defintion of $\gamma_{n,L}(\omega)$, we find, for $\qq=0$:
\begin{equation}
    T_\text{eff}    (\omega, \qq=0) \propto
    \begin{cases}
    |\omega|^{\alpha_L}  & (\text{conserved}) 
    \\
    c_1 + c_2 |\omega|^{\alpha_L - \alpha_n} & (\text{not conserved})
    \end{cases}     
\end{equation}
Thus the system with conserved angular momenta is inherently out of equilibirum, even in the low frequency limit. Moreover, this result suggestes that LRO can be established because the low energy modes are asymptotically at zero temperature.

The behavior of the non conserving system depends on the relative values of
 $\alpha_n$ and $\alpha_L$. For $\alpha_n < \alpha_L$ the effective temperature is constant at low-energies, suggesting that the low-energy physics is thermal like. This is consistent with our result  $C(\qq) \propto q^{-2}$ for this case, which matches the equlibrium correlation function. For $\alpha_n > \alpha_L$ the effective temperature shows a power-law divergence at low frequencies indicating that  the system is out of equilibrium. Since the effective temperature diverges, the system cannot sustain LRO. This is also in agreement with our finding that $C(\qq) \propto q^{-\eta}$ with $\eta >2$ in this case.

\section{Proof of non-equilibrium conditions from absence of thermal symmetry.}
\label{App:ProofNonEq}
We now give a general proof that the model defined by Eq.~\eqref{eq:Langevin} is inherently out of equilibrium. For simplicity, we focus on the case with $\alpha_L \geq 1$, where the conserved densities $L_{\alpha \beta}$ can be treated as markovian.
The proof consists of two steps: (i) we first make the system markovian by adding a fictitious bath which mimics the non-Markovian kernel $\gamma_n$, and (ii) we show that the new enlarged system is out of equilibrium by testing if its MSR action satisfies the equilibrium symmetry \cite{Janssen,janssen1992renormalized}.
We first prove step (i). To this end, we start by defining:
\begin{equation}
    \tilde{\cH}_d = \cH_d + \int_\rr \sum_\nu \omega_\nu\varphi_{\nu,\alpha}^2(\rr) + \int_\rr \sum_\nu g_\nu \varphi_{\nu,\alpha}(\rr) n_\alpha(\rr) 
\end{equation}
where we introduced the bath degrees of freedom $\varphi_{\nu,\alpha}$, with energy $\omega_\nu$ and linearly coupled to the order parameter with strength $g_\nu$.
The Markovian equation of motion for the order parameter and bath variables are:
\begin{subequations}
    \begin{align}
        \dot{n}_\alpha &= - \frac{\delta \tilde{\cH}_d}{\delta n_\alpha} + \zeta_\alpha = - \frac{\delta {\cH}_d}{\delta n_\alpha} -\sum_{\nu}g_\nu \varphi_\alpha + \zeta_\alpha, \\
        \dot{\varphi}_{\nu,\alpha} & = - \frac{\delta \tilde{\cH}_d}{\delta \varphi_{\nu,\alpha}} + \zeta_{\nu,\alpha} = - \omega_\nu \varphi_{\nu,\alpha} - g_\nu n_\alpha + \zeta_{\nu,\alpha},
    \end{align}
\end{subequations}
with $\zeta_\alpha$ and $\zeta_{\nu, \alpha}$ zero-average Gaussian white noise with correlations:
\begin{equation}
    \langle \zeta_\alpha(\rr,t) \zeta_\beta(\mathbf{0},0) \rangle =2T_n\delta_{\alpha\beta} \delta(\rr)\delta(t), \qquad
    \langle \zeta_{\nu,\alpha}(\rr,t) \zeta_{\mu,\beta}(\mathbf{0},0) \rangle =2T_n\delta_{\mu\nu}\delta_{\alpha\beta} \delta(\rr)\delta(t)
\end{equation}
Now, by solving explicitly for $\varphi_{\nu, \alpha}$, we find (assuming the initial time to be $t_0 = -\infty$):
\begin{equation}
\varphi_{\nu, \alpha}(\rr,t) = \int_{-\infty}^t dt' e^{-\omega_\nu(t-t')} \bigg[-g_\nu n_\alpha(\rr,t') + \zeta_{\nu,\alpha}(\rr,t) \bigg], 
\end{equation}
and replacing this back in the equation for $n_\alpha$, we find:
\begin{equation}
    \dot{n}_\alpha + \gamma*\dot{n}_\alpha = - \frac{\delta {\cH}_d}{\delta n_\alpha} + \tilde{\zeta}_\alpha + \zeta_\alpha 
\end{equation}
with $\tilde{\zeta}_\alpha$ an effective zero-average Gaussian noise with correlation $\langle \xi_\alpha(\rr,t) \xi_\beta(\mathbf{0},0) \rangle =T_n\delta_{\alpha\beta} \delta(\rr) [\gamma(t) + \gamma(-t) ]$, where the function $\gamma(t)$ is given by 
\begin{equation}
\label{eq:effective_equation}
    \gamma(t) = \theta(t)\sum_\nu\frac{g_\nu^2}{\omega_\nu} e^{-\omega_\nu t}.
\end{equation}
By chosing the bath parameters $g_\nu$ and $\omega_nu$ such that $\gamma(t) = \gamma \theta(t)t^{-\alpha}$, and by recalling that, for $\alpha<1$, $\gamma(t)$ dominates at low frequencies over markovian terms (here represented by $\dot{n}_\alpha$ and $\zeta_\alpha$), we find that Eq.\eqref{eq:effective_equation} is, at low frequency, exactly Eq.\eqref{eq:Langevin} in the main text (the coupling with $L_{\alpha\beta}$ can be reinstated at this point, as it plays no roles in this derivation.)
Therefore, we just proved point (i), i.e., that the following Markovian equations are equivalent to Eq.\eqref{eq:Langevin} (for $\gamma_L = \delta(t)$):
\begin{subequations}
\label{eq:markovian_equations}
\begin{align}
        \dot{n}_\alpha & =\{n_\alpha,\cH_c\}-\frac{\delta \tilde{\cH}_d}{\delta n_\alpha}+ \zeta_\alpha,\nonumber  \\
\dot{\varphi}_{\nu,\alpha} & = - \frac{\delta \tilde{\cH}_d}{\delta \varphi_{\nu,\alpha}} + \zeta_{\nu,\alpha} \\
        \dot{L}_{\alpha\beta} & =\{L_{\alpha\beta},\cH_c\}+\nabla^2\frac{\delta \tilde{\cH_d}}{\delta L_{\beta\alpha}}+ \eta_{\alpha\beta}.
    \end{align}
\end{subequations}
Notice that we replaced ${\cH_d}$ with $\tilde{\cH}_d$ in the equation for $L_{\alpha\beta}$, since the coupling with between the bath and the order parameter is independent of $L_{\alpha\beta}$. 

We can now move to the the proof of point (ii). To this end, it is convenient to write the MSR action of Eqs.\eqref{eq:markovian_equations}, which reads~\cite{Tauber}:
\begin{multline}
\label{eq:MSR}
  S=\int_{\rr,t}\bigg\{
   \tilde{n}_{\al} \Big[\dot{n}_{\al}-
   \{n_\alpha,\cH_c\}+\frac{\delta \tilde{\cH}_d}{\delta n_\alpha}
   -T_n \tilde{n}_{\al}\Big]
   +
   \sum_\nu\tilde{\varphi}_{\nu, \alpha} \Big[\dot{\varphi}_{\nu,\al}+\frac{\delta \tilde{\cH}_d}{\delta \varphi_{\nu,\alpha}}
   -T_n \tilde{\varphi}_{\nu,\al}\Big]
   \\
+\tilde{L}_{\bb\al}\Big[\dot{L}_{\al\bb}-\{L_{\alpha\beta},\cH_c\}-\nabla^2\frac{\delta \tilde{\cH}_d}{\delta L_{\beta\alpha}}
   +T_L\nabla^2\tilde{L}_{\al\bb}\Big]
    \!\bigg\}\!.
\end{multline}
For clarity, we omit the $(\rr,t)$ dependence of the fields.
The presence of thermal equilibrium can be shown by the presence of a thermal symmetry of the MSR action, i.e., if the action~\eqref{eq:MSR} is symmetric under the following transformation $\mathcal{T}$~\cite{Janssen,janssen1992renormalized}
\begin{equation}
\label{eq:thermal_symmetry}
\begin{aligned}[c]
   &\mathcal{T}n_{\al}(t)= n_{\al}(-t),\nonumber\\
    &\mathcal{T}\tilde{n}_{\al}(t)=-\left[\tilde{n}_{\al}(-t)-\frac{1}{T}\frac{\delta\cH}{\delta n_{\al}(-t)}\right]\!,
\end{aligned}
\,
\begin{aligned}[c]
 &\mathcal{T}L_{\al\bb}(t)=- L_{\al\bb}(-t),\nonumber\\
    &\mathcal{T}\tilde{L}_{\al\bb}(t)=\left[\tilde{L}_{\al\bb}(-t)-\frac{1}{T}\frac{\delta\cH}{\delta L_{\bb\al}(-t)}\right]\!,
\end{aligned}
\,
\begin{aligned}[c]
   &\mathcal{T}\varphi_{\nu,\al}(t)= \varphi_{\nu,\al}(-t),\nonumber\\
    &\mathcal{T}\tilde{\varphi}_{\nu,\al}(t)=-\left[\tilde{\varphi}_{\nu,\al}(-t)-\frac{1}{T}\frac{\delta\cH}{\delta \varphi_{\nu,\al}(-t)}\right]\!,
\end{aligned}
\end{equation}
where for some $T$ and some function $\cH$. If the transformation $\mathcal{T}$ is a symmetry, then $T$ is the temperature of the equilibrium system, and $\cH$ is the Hamiltonian functional of the equilibrium system (whose probability distribution is the Boltzmann weight $\propto \exp[-\cH/T]$).

By substituting the transformation \eqref{eq:thermal_symmetry} into the action \eqref{eq:MSR}, one finds that the following conditions need to be met in order for $\mathcal{T}$ to be a symmetry of the action (and therefore for the system to be at thermal equilibrium):
\begin{equation}
   T_n = T_L, \qquad \{\tilde{\cH}_d, \cH_c\}=0 \quad  (\text{or, equivalently,}\quad  \tilde{\cH}_d \propto \cH_c ). 
\end{equation}
Since $\tilde{\cH}_d$ and $\cH_c$ cannot be proportional to each other, because $\tilde{\cH}_d$ contains the bath variables, while $\cH_c$ does not, we conclude that the system is out of thermal equilibrium. Accordingly, Eqs.\eqref{eq:Langevin} describe a nonequilibrium system.

\section{Comparison between scaling analysis and numerics from Bogoliubov equations}
\label{app:Plot}

\begin{figure*}[h]
	\includegraphics[width=1
	\linewidth]{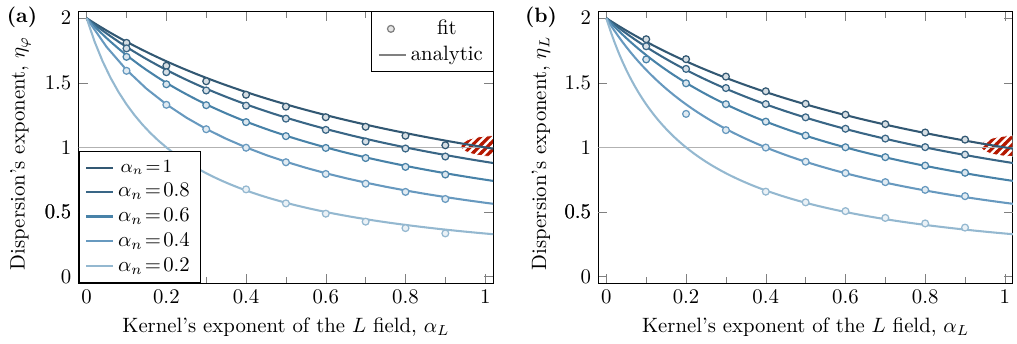}
 \caption{Extracted correlation exponent $\eta_i$ at small momenta $\qq$ as functions of $\alpha_L$ for various values of $\alpha_{n}$. (a) For $i=\varphi$,
 the correlation function for the $\varphi$ fields is parametrized as  $C_{\varphi\varphi} \approx 1/A_{\varphi}|\qq|^{\eta_{\varphi}}$. (b) For $i=L$,
 the correlation function for the $L$ fields is parametrized as  $C_{LL} \approx A_L |\qq|^{\eta_L}$.
 The solid lines show the analytically obtained estimate of the scaling in the IR. The gray line indicates the value above which order in $d=1$ can be sustained.
 The red shaded region indicates where the Markovian limit is recovered. However, 
$\eta$ does not reach the typical equilibrium scaling of 2 for the correlation function of the $\varphi$ fields. Instead, scaling is recovered as the prefactor of $\qq^{2\alpha_ n/(\alpha_n+\alpha_L)}$ vanishes, making the subleading 
$q^2$ term dominant, as explained in more detail in the main text.}
 \label{Fig:DispersionPlot}
\end{figure*}

\section{Special case $\alpha_{n,L}=0$}
\label{app:alphazero}

We consider here the special case when either $\alpha_n$ or $\alpha_L$ vanish. They key observation is that, in this case, the time derivative in the stochastic equations~\eqref{eq:Langevin} can be written as follows, using Eq.~\eqref{eq:gamma_expression} (considering $n$ as an example):
\begin{equation}
    \int_{-\infty}^{+\infty} dt' \gamma_{n}(t-t') \dot{n}_\alpha(t') = \gamma_n\int_{-\infty}^{t} dt'  \dot{n}_\alpha(t') = \gamma_n n_\alpha(t),
\end{equation}
indicating that the time-derivative term is effectively replaced by a mass-like term. We consider now the impact of this in the two different cases $\alpha_n=0$ and $\alpha_L=0$. We consider the linearize equation in the ordered phase.\\

\emph{Case $\alpha_n=0$} --- The equations in this case read:
\begin{equation}
   \!\!\!\! \begin{pmatrix}
        \gamma_n + D_nq^2 & g\bar{n}D_L \\
        -g\bar{n}D_nq^2 & -i\omega\gamma_L(\omega) + D_Lq^2
    \end{pmatrix}\!\!
    \begin{pmatrix}
        \varphi_\alpha(\qq,\omega) \\
        \ell_{\alpha}(\qq,\omega)
    \end{pmatrix}
    \!=\!
    \begin{pmatrix}
        \xi_\alpha(\qq,\omega) \\
        \eta_{1\alpha}(\qq,\omega)
    \end{pmatrix},
\end{equation}
which show that the Goldstone mode $\varphi_\alpha$ has become a massive field (with mass $\gamma_n$). The coupling with the massless field $L_{1\alpha}$ cannot remove the Goldstone mass, as it only generates frequency or momentum-dependent terms. This implies that no infrared divergence can take place, and therefore the Goldstone modes cannot destabilize LRO.  \\

\emph{Case $\alpha_L = 0$} --- The equations now read:
\begin{equation}
   \!\!\!\! \begin{pmatrix}
        -i\omega\gamma_n(\omega) + D_nq^2 & g\bar{n}D_L \\
        -g\bar{n}D_nq^2 & \gamma_L + D_Lq^2
    \end{pmatrix}\!\!
    \begin{pmatrix}
        \varphi_\alpha(\qq,\omega) \\
        \ell_{\alpha}(\qq,\omega)
    \end{pmatrix}
    \!=\!
    \begin{pmatrix}
        \xi_\alpha(\qq,\omega) \\
        \eta_{1\alpha}(\qq,\omega)
    \end{pmatrix},
\end{equation}
In this case the conserved density $L_{1\alpha}$ becomes massive with mass $\gamma_L$. The coupling of $L_{1\alpha}$ to the Goldstone mode $\varphi_\alpha$ only renormalizes the diffusion term $\propto q^2$. Accordingly, the dynamics of the Goldstone modes is described by the non-Markovian model A which, as discussed in the main text, is in thermal equilibrium. Accordingly, the Goldstone modes will generate infra-red divergences for $d\leq 2$ and destroy LRO, according to the Mermin-Wagner theorem.

\end{document}